# Physically Consistent Machine Learning for Melting Temperature Prediction of Refractory High-Entropy Alloys

Mohd Hasnain, Department of Applied Physics and Materials Engineering
National Institute of Technology Patna
Mail: mohdh.ug23.ph@nitp.ac.in

**Abstract :**

Predicting the melting temperature (Tm) of multi-component and high-entropy alloys (HEAs) is critical for high-temperature applications but computationally expensive using traditional CALPHAD or DFT methods. In this work, we develop a gradient-boosted decision tree (XGBoost) model to predict Tm for complex alloys based on elemental properties. To ensure physical consistency, we address the issue of data leakage by excluding temperature-dependent thermodynamic descriptors (such as Gibbs free energy of mixing) and instead rely on physically motivated elemental features. The optimized model achieves a coefficient of determination ($R^2$) of 0.948 and a Mean Squared Error (MSE) of 9928 which is about 5% relative error for HEAs on a validation set of approximately 1300 compositions. Crucially, we validate the model using the Valence Electron Concentration (VEC) rule. Without explicit constraints during training, the model successfully captures the known stability transition between BCC and FCC phases at a VEC of approximately 6.87. These results demonstrate that data-driven models, when properly feature-engineered, can capture fundamental metallurgical principles for rapid alloy screening.

## 1. Introduction

The melting temperature (Tm) is a fundamental thermodynamic property that governs the high-temperature applicability of metallic materials. Conventionally, structural alloys are based on a single principal element (e.g., Fe, Al, or Ti) with minor alloying additions. In contrast, High-Entropy Alloys (HEAs) represent a paradigm shift in alloy design, consisting of five or more principal elements in near-equiatomic concentrations **[1, 2]**. These multi-principal element alloys are hypothesized to be stabilized by high configurational entropy, which suppresses the formation of brittle intermetallic compounds in favor of simple solid-solution phases such as FCC or BCC structures **[1]**. Within this class, **Refractory High-Entropy Alloys (RHEAs)**, composed primarily of elements like W, Nb, Mo, Ta, and V, have attracted significant attention for aerospace and energy applications due to their exceptional high-temperature strength and thermal stability **[3].**

In the context of these multi-component systems, predicting Tm is particularly challenging due to complex chemical interactions, severe lattice distortion, and the vastness of the compositional space. Traditional physics-based approaches such as CALPHAD or first-principles calculations (DFT) become computationally expensive or unreliable when extended to these large, disordered compositional spaces. Machine learning (ML) offers a promising alternative by learning composition-property relationships directly from data. However, purely data-driven models risk learning spurious correlations unless their predictions are validated against established physical principles.

In this work, we combine interpretable machine learning with electronic-structure-based validation to ensure physically consistent melting temperature predictions.

## 2. Dataset and Feature Engineering

The dataset consists of experimentally reported melting temperatures of multi-component alloys containing transition and refractory elements. The dataset used in this study was adapted from **Machaka et al. [4]**, consisting of approximately 1300 multi-component alloy compositions with experimentally reported properties. While the original dataset was curated for phase classification (BCC/FCC/Amorphous), it includes verified melting temperatures (Tm) and compositional features which serve as the ground truth for our regression analysis. Each alloy composition is represented using a set of physically motivated descriptors derived from elemental properties.

Key descriptors include:

- Valence Electron Concentration (VEC)
- Thermodynamic parameters such as enthalpy of mixing and atomic size mismatch
- Composition-weighted elemental properties

These features are designed to encode both electronic and thermodynamic information relevant to phase stability and bonding strength.

Problem of Data Leakage:

During preliminary modeling, certain thermodynamic descriptors were found to introduce data leakage. In particular, the Gibbs free energy of mixing ($\Delta G_{mix}$), when computed at the melting temperature, becomes linearly dependent on the target variable through the relation $\Delta G_{mix} = \Delta H_{mix} - T_m \Delta S_{mix}$

Since $\Delta S_{mix}$ varies weakly across compositions, this can artificially inflate predictive performance. To avoid this issue, such features were excluded, ensuring that the model learns physically meaningful correlations rather than trivial functional dependencies.

## 3. Machine Learning Model

A gradient-boosted decision tree model (XGBoost) was employed to learn the non-linear relationship between alloy descriptors and melting temperature. The model was trained using cross-validation, and hyperparameters were optimized via randomized and grid-based search to minimize prediction error.

Model performance was evaluated using standard regression metrics, including the coefficient of determination ($R^2$) and mean squared error (MSE). The optimized model achieved high predictive accuracy, indicating its ability to capture complex composition–property relationships.

| Model | Coefficient of Determination($R^2$) | Mean Squared Error (MSE) |
|---|---|---|
| Linear Regression (Baseline) | -35.071155362825245 | 6921263.106562936 |
| RandomForest | 0.9335757462485741 | 12745.356566663355 |
| XGBoost (hyperparameter optimized) | 0.9483030675540977 | 9928.789822069666 |

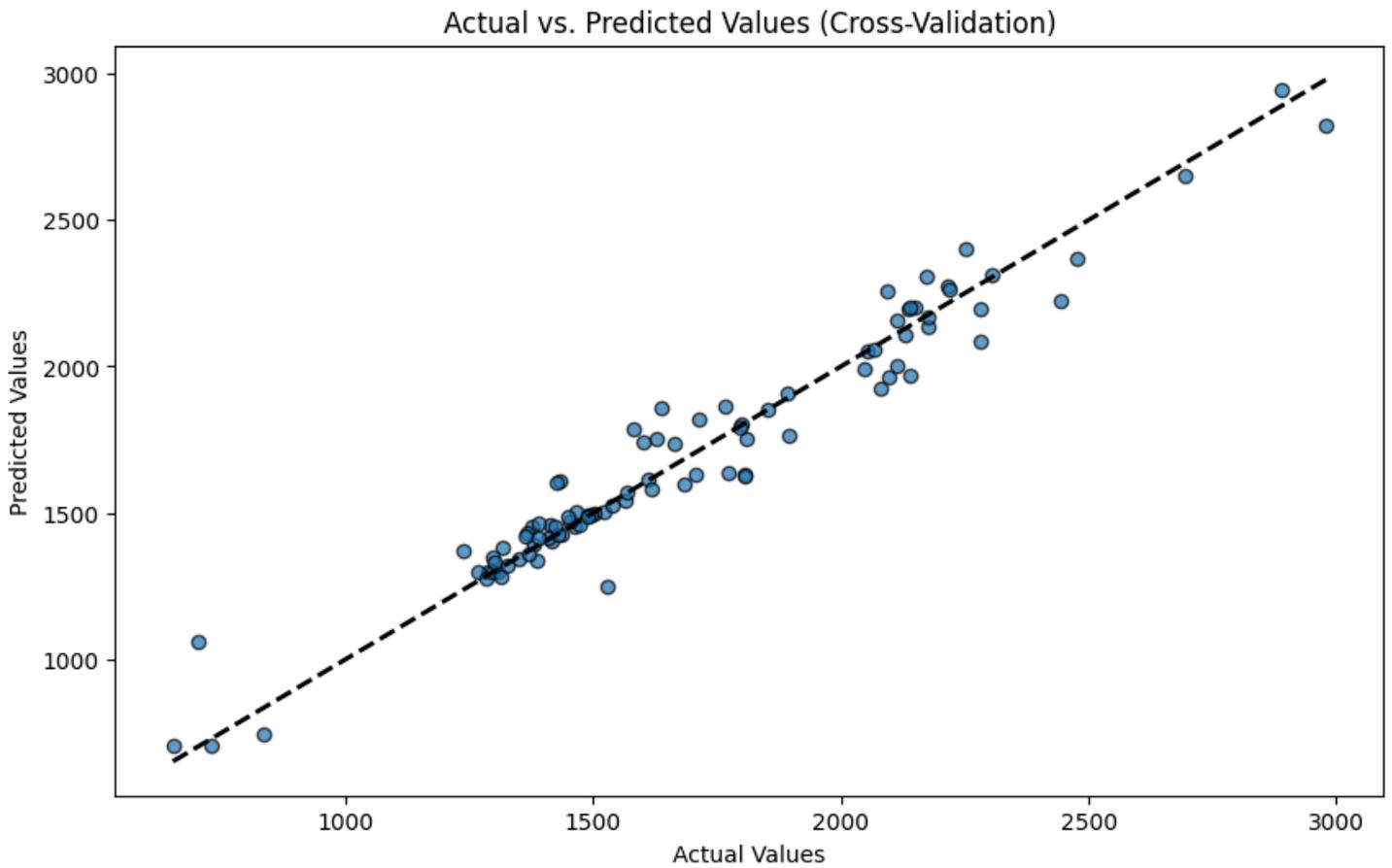

*Figure 1: Predicted melting temperatures vs the actual values of melting temperature for HEA.*

## 4. Results and Model Interpretability

The trained model demonstrates strong agreement between predicted and experimental melting temperatures. To interpret the model behavior, feature importance analysis and SHAP (SHapley Additive exPlanations) were employed. These analyses reveal that electronic and thermodynamic descriptors, particularly those related to valence electron concentration and refractory element content, play a dominant role in determining Tm.

    The model exhibits strong alignment with metallurgical principles. Refractory Elements (Nb, Ta, Ti) are identified as the primary positive contibutors to melting temperature  (positive SHAP values for high concentrations ), while low-melting elements (Cu, Mn, Mg, Zn) are correctly identified as melting point depressants. Additionally, VEC (Valence Electron Concentration) appears as a top tier descriptor (Red dots shows positive impact), suggesting the model has captured the electronic stability criteria often used to predict high-stability BCC Phases in High Entropy Alloys.

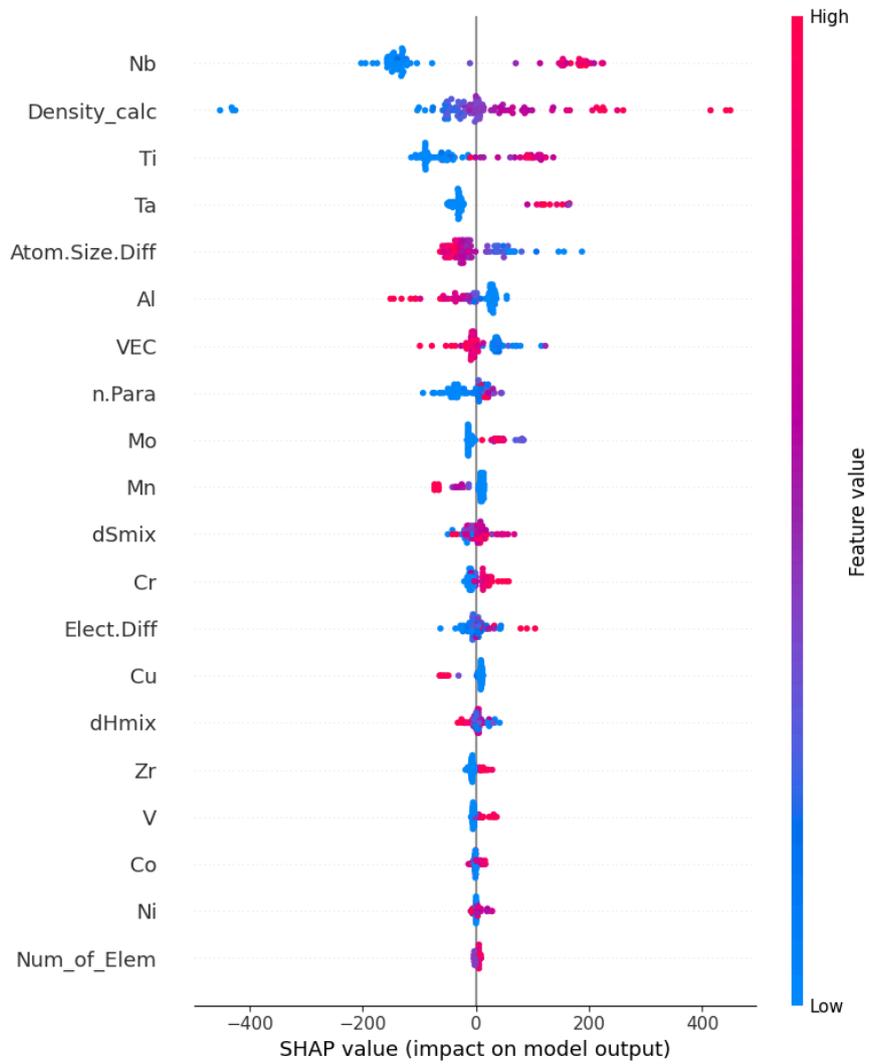

*Figure 2: SHAP (SHapley Additive exPlanations) summary plot illustrating global feature importance. Features are ranked by mean absolute SHAP value, with the most influential features at the top. Each dot represents a single alloy composition; color indicates the feature value (red = high, blue = low). The horizontal position shows the impact on the predicted melting temperature. The model correctly identifies refractory elements (Nb, Ti, Ta) as strong positive contributors to thermal stability, while low-melting elements (Cu, Mn, Zn) act as melting point depressants.*

## 5. Physical Validation Using Valence Electron Concentration

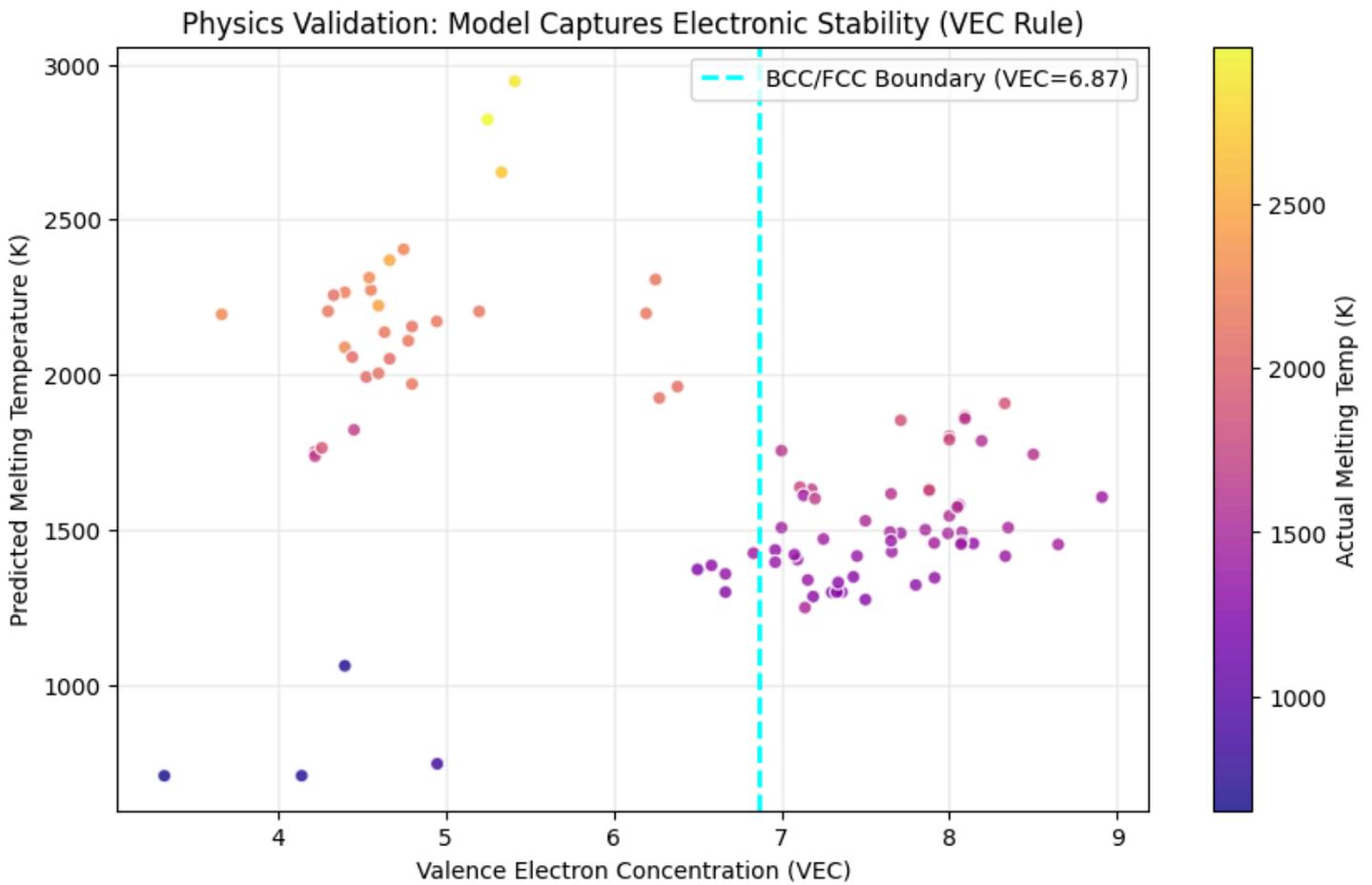

To assess whether the model predictions are physically meaningful, predicted melting temperatures were analyzed as a function of Valence Electron Concentration (VEC).Here, physical consistency refers to agreement with established electronic-structure trends, rather than hard constraints during training. A scatter plot of predicted Tm versus VEC reveals a clear separation of data points across a critical VEC value of approximately 6.87.

Alloys with VEC values below this threshold predominantly exhibit higher predicted melting temperatures, while alloys with higher VEC values cluster at lower melting temperatures. This transition aligns with the well-known electronic stability criterion in transition-metal alloys, where lower VEC values favor body-centered cubic (BCC) structures typically associated with refractory elements and high melting points, whereas higher VEC values favor face-centered cubic (FCC) structures with comparatively lower melting temperatures.

Importantly, this behavior emerges naturally from the model predictions and was not explicitly enforced during training. The agreement between model output and established electronic-structure trends provides strong evidence that the model has learned physically relevant representations rather than relying solely on statistical correlations.

*Figure 3: Validation of predicted melting temperatures against the Valence Electron Concentration (VEC) rule. The vertical line at VEC= 6.87 marks the BCC/FCC stability boundary.*

# 6. Conclusions and Future Work

In this work, a physics-informed machine learning framework was developed to predict the melting temperature of complex alloys. Beyond achieving high predictive accuracy, the model was validated against known electronic stability criteria, demonstrating consistency with fundamental alloy physics.

Future work will focus on extending this approach toward inverse alloy design, integrating uncertainty quantification, and coupling ML predictions with first-principles calculations for further validation. Such extensions would enable systematic discovery of novel high-temperature materials and strengthen the integration of data-driven methods with physical metallurgy.

**Data and Code Availability** The complete source code, including the data preprocessing pipeline and the optimized XGBoost model, is available in the following repository:

https://github.com/h328j2/HEA-melting-temperature-pridiction